\documentclass[dvips]{article}
\usepackage{icrctc07}

\title{Gamma-ray emission associated with the Cluster-scale AGN Outbursts in the Hydra A system}
\shorttitle{Gamma rays from Hydra A}
\authors{W. Domainko$^{1}$, J. A. Hinton$^{2}$, E. C. D. Pope$^{2}$}
\shortauthors{W. Domainko et al}
\afiliations{$^1$Max-Planck-Institut f\"ur Kernphysik, Heidelberg, Germany\\ $^2$School of Physics \& Astronomy University of Leeds, UK}
\email{wilfried.domainko@mpi-hd.mpg.de}

\abstract{Recent observations have revealed the existence of an enormously
energetic $>$ 10$^{61}$ erg AGN outburst in the Hydra A cluster of galaxies.
This outburst has produced cavities in the
intra-cluster medium, apparently supported by pressure from
cosmic rays. Here we argue that if these cavities are filled with $>$ GeV particles, 
these particles are very
likely protons and nuclei. For a plausible
spatial distribution of the target gas, based on observations and
hydrodynamical simulations, we show that the $\pi^0$-decay gamma-rays
from these cosmic-rays may be detectable with the H.E.S.S.experiment.}

\begin{document}
\maketitle

\section{Introduction}

In several galaxy clusters interaction between AGN outbursts and the intra-cluster medium (ICM) can be seen. High resolution X-ray observations have revealed the presence of bubbles, cavities and weak shocks in the ICM driven by the activity of the nucleus in the central galaxy (e.g. \cite{boehringer93}, \cite{blanton01}, \cite{schindler01}, \cite{mcnamara01}, \cite{fabian03}, \cite{choi04}, \cite{birzan04}). These structures in the X-ray emitting gas are often associated with radio lobes, thus indicating the presence of relativistic electrons \cite{owen00}, \cite{fabian02}, \cite{gitti06}. 

In some cases the energy related to AGN outbursts can exceed $10^{61}$ erg \cite{mcnamara05}, \cite{nulsen05a}, \cite{nulsen05b}, \cite{wise07}, \cite{gitti07}. In such cases the feedback from the AGN can be seen even at large distances from the cluster center and therefore this phenomenon is often called \textit{cluster scale AGN outbursts}. Also in these systems the X-ray surface brightness depressions are coincident with radio lobes thus showing the presence of non-thermal electrons \cite{lane04}, \cite{cohen05}. The typical timescale of cluster scale AGN outbursts is about 10$^8$ years. One system featuring such an outburst is Hydra A.

High energy particles which are injected by AGNs in clusters of galaxies can produce gamma rays through various processes (see \cite{blasi07} for a recent review). Inelastic collisions between cosmic ray protons and target nuclei in form of the thermal intra-cluster medium (ICM) can result in emission of gamma rays through $\pi^0$ decay \cite{dennison80}, \cite{voelk96}. Cosmic microwave background (CMB) photons can be up scattered by high energy electrons to gamma-ray energies in inverse compton processes \cite{atoyan00},\cite{gabici03}, \cite{gabici04}.

\section{Model assumption}

In this contribution we investigate a scenario were the bubbles in the X-ray gas of Hydra A are expanded by hadronic cosmic rays and produce gamma rays due to inelastic collisions with nuclei of the thermal plasma and subsequent $\pi^0$ decay. For a more extended model description see \cite{hinton07}.

\textit{Energetics:} The total energetics of AGN outbursts in a galaxy cluster environment can be estimated through the work which is done on the ICM. AGN jets will inflate bubbles with non-thermal particles in the ICM and the total energy related with the outburst is then constrained by the PdV work which is done by expansion of the bubble. In case of Hydra A the total energy of the outburst is close to 10$^{61}$ erg.

\textit{Composition of the non-thermal particles inside the lobes:} The age of the outburst in the Hydra A system was estimated to be in the order of 10$^8$ years. The energy loss time scale of $>$ GeV electrons in typical cluster environments with magnetic fields in the $\mu G$ range is about 10$^6$ years. Therefore high energy electrons can not support the expansion of the observed bubbles in the thermal ICM over their entire evolution. On basis of this considerations we favor hadronic cosmic rays over leptonic cosmic rays to do the work on the ICM which is evident in the spatial distribution of the thermal X-ray emitting gas. But it hast to be said that alternatives to this model in form of low energy electrons or magnetic fields still exist. 

\textit{Density of the target material:} The density of target material in galaxy cluster can in principle be estimated with X-ray measurements. In case of bubbles embedded in the ICM the situation is more complicated since the 3d location of the lobes created by the AGN outburst is not known. From X-ray data, only an upper limit can be placed for the density inside the bubbles with respect to the density of the surrounding ICM. In case of the Hydra A system the ratio between the density of thermal plasma inside and outside the radio lobes is less than 0.3 \cite{wise07}. But gas which could be present in the bubbles and does not emit in X-rays would escape detection and the resulting density inside the cavities could be larger then indicated by the observations.

\section{Gamma ray emission}

\begin{figure*}
\begin{center}
\includegraphics [width=0.85\textwidth]{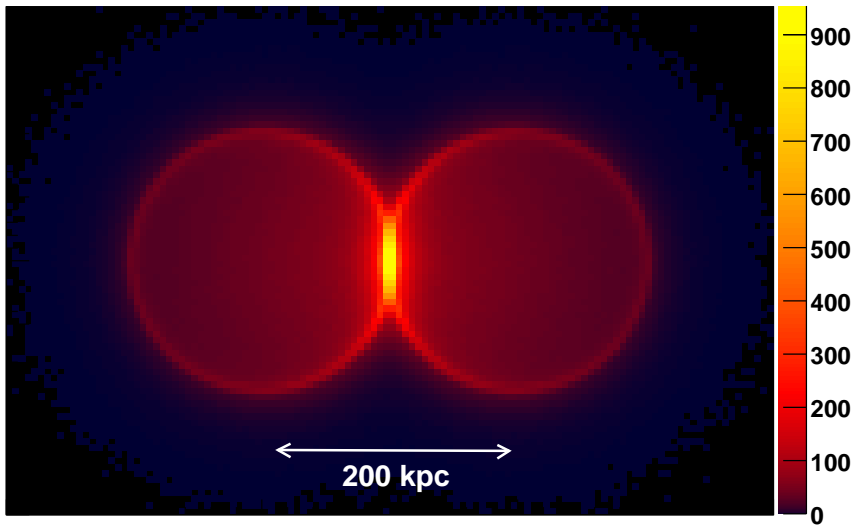}
\end{center}
\caption{Surface brightness of the proposed gamma ray emission of the bubbles in the Hydra A system. Hadronic cosmic rays which diffused in the high density walls of the cavities lead to an enhanced gamma ray emission. Surface brightness in arbitrary units.}\label{gamma}
\end{figure*}

In the proposed scenario gamma rays are produced through inelastic collisions between cosmic ray protons and nuclei from the thermal ICM. We used the SIBYLL hadronic interaction model \cite{fletcher94}, \cite{kelner06} to calculate the rate of proton-proton collisions. To derive the gamma-ray luminosity of clusters hosting large scale AGN outbursts the distribution of cosmic rays and the distribution of the target material has to be known. We assume that the cosmic rays diffuse out of the bubbles according the model by \cite{voelk96} and we further assume that the rims of the lobes have a four times larger density than the surrounding gas (see Fig. \ref{gamma}). The gamma-ray brightness is then calculated for different densities of target material inside the bubbles. In Fig. \ref{SED} the situation is shown for bubbles filled with cold gas which is not seen in X-rays and for a situation were the bubbles do not contain any thermal gas.

\section{Prospect for the detectability with present and future gamma ray telescopes}

\begin{figure*}
\begin{center}
\includegraphics [width=0.90\textwidth]{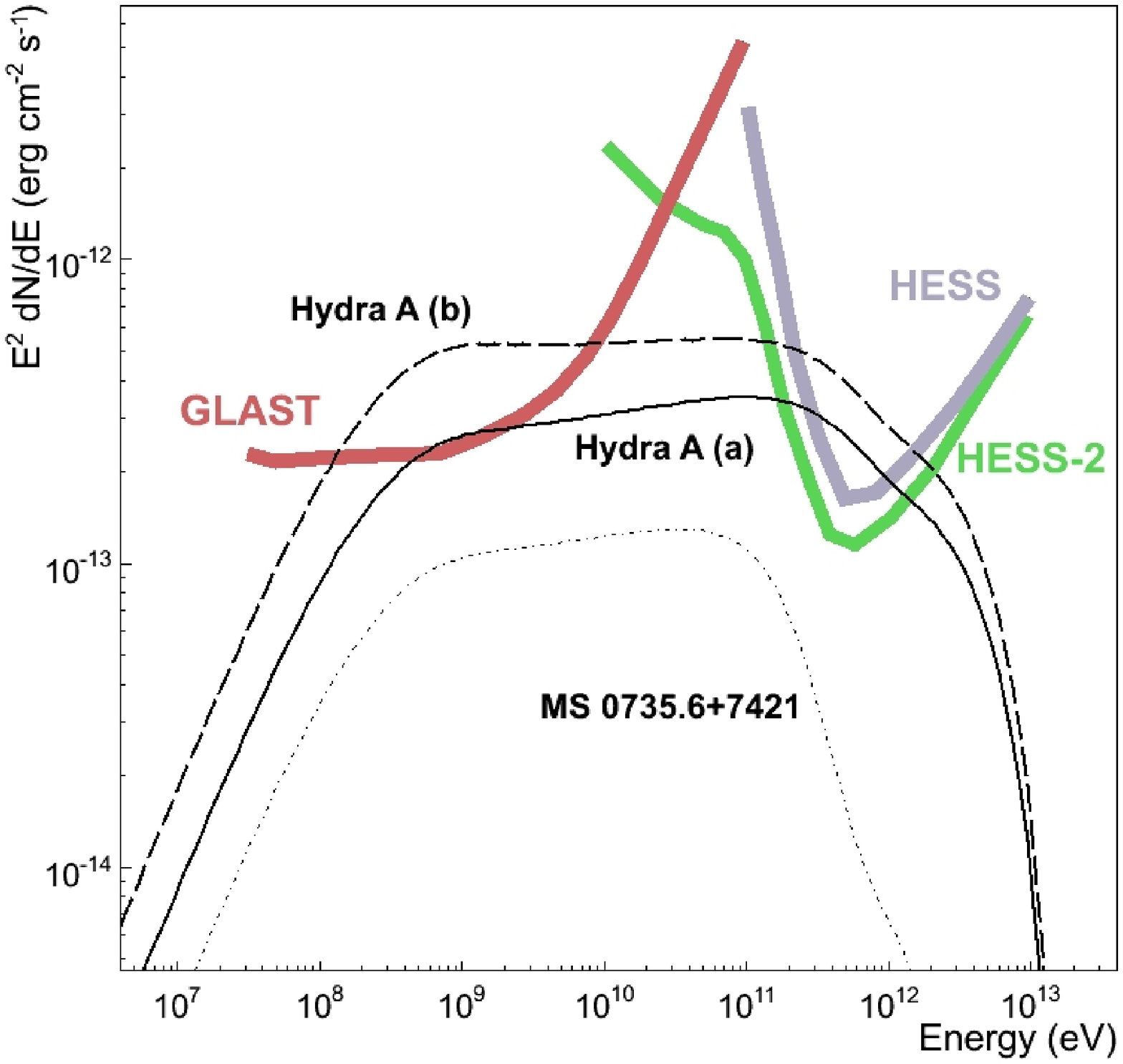}
\end{center}
\caption{Predicted gamma-ray brightness of the Hydra A system assuming that the observed bubbles are entirely supported by hadronic cosmic rays. Case (a) shows the expectation for a scenario were the bubbles do not contain thermal plasma and case (b) corresponds to bubbles filled with cold gas which is not seen in X-rays. The emission may be detectable with the H.E.S.S. experiment and the GLAST satellite.}\label{SED}
\end{figure*}

Gamma ray astronomy is in a phase of rapid development. Several Imaging Atmospheric Cherenkov Telescopes (IACTs) are observing in the 100 GeV - 100 TeV range: HESS~\cite{hinton04},
MAGIC~\cite{lorenz04} and VERITAS~\cite{krennrich04}. Additionally to these ground based instruments GLAST \cite{thompson04} will be launched early next year and will then provide data in the 10 MeV - 100 GeV band. We find that in the framework of the presented model the gamma-ray brightness of the Hydra A system may indeed be close to the sensitivity limit of these operating or upcoming instruments (see Fig. \ref{SED}).

\section{Acknowledgments}

We would like to thank Felix Aharonian, Heinz V\"olk, Olaf Reimer and
Werner Hofmann for very useful discussions. JAH is supported by a PPARC Advanced Fellowship.

\end{document}